\documentclass[aps,prb,twocolumn,groupedaddress,showpacs,floatfix]{revtex4}
\usepackage{epsfig}
\usepackage{epsf} 
\begin{document}
%
\title{Ab initio study of the $\beta$-tin$\to${\it Imma}$\to$sh phase
  transitions in 
  silicon and germanium} 

\author{Katalin Ga\'al-Nagy, Pasquale Pavone, and Dieter Strauch}

\affiliation{Institut f{\"u}r theoretische Physik, Universit{\"a}t
  Regensburg, D-93040 Regensburg, Germany}

\date{\today}
%
\begin{abstract}
We have investigated the structural sequence of the high-pressure
phases of silicon and germanium. We have focussed on the cd $\to$
$\beta$-tin $\to$ {\it Imma} $\to$ sh phase transitions. We have
used the plane-wave pseudopotential approach to the
density-functional theory implemented within the Vienna {\it
ab-initio} simulation package (VASP). We have determined the
equilibrium properties of each structure and the values of the
critical parameters including a hysteresis effect at the phase
transitions. The order of the phase transitions has been obtained
alternatively 
from the pressure dependence of the enthalpy and of the internal
structure parameters. The commonly used tangent construction is
shown to be very unreliable. Our calculations
identify a first-order phase transition from the cd to the
$\beta$-tin and from the {\it Imma} to the sh phase, and they
indicate the possibility of a second-order phase-transition from the
$\beta$-tin to the {\it Imma} phase. Finally, we have derived the
enthalpy barriers between the phases.
\end{abstract}
\pacs{61.50.Ks, 64.70.Kb, 71.15.Nc, 71.20.Mq, 81.30.Dz, 81.40.Vw}
\maketitle

%
\section{Introduction}\label{Intro}
Although the structural sequence of the high-pressure phase
transitions in Si and Ge has been considered to be well known since
two decades, a new high-pressure phase was found
experimentally between the body-centered tetragonal structure with
two basis atoms (which will be referred to the following as
$\beta$-tin or BCT phase) and the simple-hexagonal (sh) structure in
Si\cite{McMahon1993,McMahon1994} and
Ge.\cite{Nelmes1996,McMahon1996} After its space group this
intermediate phase was called {\it Imma} phase and corresponds to a
body-centered orthorhombic Bravais lattice with two atoms in the
unit cell (which will be called in the following BCO). Thus, the
experimental structural sequence is found to be cd $\to$ $\beta$-tin
$\to$ {\it Imma} $\to$ sh $\to\ldots$, where cd indicates the
cubic-diamond structure. A BCO structure was first theoretically
proposed for Si by Needs and Martin\cite{Needs1984} about twenty
years ago, whereas Lewis and Cohen\cite{Lewis1994} predicted exactly
the {\it Imma} phase for Ge a few years before the experimental
confirmation. The first theoretical investigations for Si and Ge
indicated for both, the $\beta$-tin$\to${\it Imma} and the {\it
Imma}$\to$sh case, a continuous second-order phase
transition.\cite{Lewis1994,Lewis1993} At variance, the experiments
show first-order phase transitions with a discontinuity of the
volume in both substances and a hysteresis effect in
Si.\cite{McMahon1994} For Ge the first-order transition was detected
only for the $\beta$-tin$\to${\it Imma} transition,\cite{Nelmes1996}
because the {\it Imma}$\to$sh transition was expected to be beyond
the experimentally accessible pressure range. The transition
pressures in Ge are expected to be higher than the corresponding
ones in Si, because of the strong repulsive character of the atomic
potential in Ge due to the presence of $d$ electrons in the
core.\cite{Yin1982} From the theoretical point of view, recent
total-energy calculations show two first-order phase transitions
$\beta$-tin$\to${\it Imma} and {\it Imma}$\to$sh both for
Si\cite{Christensen1999} and Ge.\cite{Ribeiro2000} A second-order
phase transition was found for the $\beta$-tin$\to${\it Imma} case
from elastic instabilities.\cite{Hebbache2001}

First information on the order of the phase transition can be given
by group-theoretical arguments:\cite{Boccara1968} If the order of
the point group of one phase is one half of the order of the point
group of the other phase, then the phase transition can be of second
order. If the point group of one phase is one third of the order of
the point group of the other phase, then the phase transition must
be of first order. The order of the point group $m\bar{3}m$ of the
cd structure (space group $O_h^7$, $Fd\bar{3}m$) is 48. Since the
order of the point group $4/mmm$ of the $\beta$-tin phase (space
group $D_{4h}^{19}$, $I4_1/amd$ ) is 16, the cd$\to$ $\beta$-tin
transition has to be of first order. The order of the point group
$mmm$ of the {\it Imma} phase (space group $D_{2h}^{28}$, $Imma$) is
8, thus the $\beta$-tin$\to${\it Imma} transition can be a
second-order one. Furthermore, the order of the point group $6/mmm$
of the sh structure (space group $D^1_{6h}$, $P6/mmm$) is 24, the
sh$\to${\it Imma} and also the {\it Imma}$\to$sh transitions are of
first order. Following the group-theoretical analysis none of the
available total-energy calculations have described correctly the
order for the phase transitions of the whole sequence, neither for Si
nor for Ge.

This paper is organised as follows: In Section~\ref{method} we give
a short overview of the methods on which our calculations are
based. In Section~\ref{parameters} we report the results for the
equilibrium parameters of these phases and compare them with the
experimental volume dependence. Transition pressure ranges are
presented in Section~\ref{pressure}. The order of the phase
transitions is analysed in Section~\ref{order}, and the enthalpy
barriers between the phases are determined in
Section~\ref{barriers}. All results are summed up and discussed in
Section~\ref{discussion}. The conclusions are summarised in
Section~\ref{conclusion}.
%
%
\section{Method}\label{method}
Our calculations have been performed using the plane-wave
pseudopotential approach to the density-functional theory
(DFT)\cite{Hohenberg1964,Kohn1965} implemented within the Vienna
{\it ab-initio} simulation package (VASP).\cite{Kresse1996} For the
exchange-correlation potential we have used the generalised-gradient
approximation (GGA) by Perdew and Wang\cite{Perdew1992} for Si and
Ge, and, because of poorer results, the local-density approximation
(LDA)\cite{Perdew1981,Ceperley1980} for Si only.  The interaction
with the ion cores has been described by ultrasoft
pseudopotentials.\cite{Vanderbilt1985,Kresse1994} The wave functions
have been expanded in terms of plane waves up to a kinetic-energy cutoff of
270~eV (410~eV) for Si (Ge). This choice provided an error smaller
than 0.5~kbar (0.2~kbar) for Si (Ge) to the pressure according to
the Pulay stress.\cite{Pulay1980} The Pulay stress arises from using
an incomplete plane-wave basis set and causes errors in the stress
tensor and, correspondingly, in the pressure. Convergence also
required a 18$\times$18$\times$18 (24$\times$24$\times$24) mesh of
Monkhorst-Pack points\cite{Monkhorst1976} which yielded 864 (1962)
{\bf k}-points in the irreducible wedge of the Brillouin zone for Si
(Ge). All high-pressure phases are metallic. Thus, we have used a
Methfessel-Paxton smearing\cite{Methfessel1989} with a width of
0.2~eV. In order to determine the total energy for a given volume or
structure we have minimised the energy with a conjugate-gradient
algorithm.\cite{Press1986,Teter1989,Bylander1990} The equilibrium
properties have been obtained by fitting the values of the energy as
a function of the volume alternatively to the Vinet\cite{Vinet1986}
and to the Murnaghan\cite{Murnaghan1944} equation of state.
%
%
\section{Structural relaxation and equilibrium
  properties}\label{parameters} 
In this section we present our results for the volume dependence of
the structural parameters and compare them with the experimental
ones. The structural parameters are more useful indicators of phase
transitions than the often-used tangent construction to the
equation-of-state curves, the difficulties of which will be
illuminated. 

We have focussed our attention to the structural sequence
$\beta$-tin$\to${\it Imma}$\to$sh. The calculations for the cd phase
have been performed just for the sake of comparison. The most
general structure which is compatible with all four phases is the
body-centered orthorhombic cell with a basis of two atoms at (0,0,0)
and (0,$0.5b$,$\Delta\,c$) (BCO structure). The other structures are
special cases: For the BCT structure ($\beta$-tin) one has the
lattice constants $a=b$ and the internal parameter $\Delta=0.25$;
for the sh structure one has $b=\sqrt{3}c$ and $\Delta=0.5$; for the
cd phase one has $b=a$, $c=\sqrt{2}a$, and $\Delta=0.25$. The pure
{\it Imma} phase is realised as a BCO structure with
$a\not=b\not=c\not=a$ and $0.25<\Delta<0.5$. The minimization of the
total energy has been performed using the following procedure: For
all phases we have started from the BCO structure and minimised the
energy as a function of the structural parameters with the
constraints appropriate to the various phases. In this way we have
avoided an energy offset which arises from using different
cells. For example, in the case of the cd structure we have imposed,
according to the previous considerations, $b=a$, $c=\sqrt{2}a$, and
$\Delta=0.25$, and then we have minimised the total energy as a
function of $a$. In the special case of the sh structure, we could
keep only the value of $\Delta$ fixed at 0.5, and we have relaxed
the value of $b/c$ (and also $c/a$) by minimization of the energy;
the structure which is obtained after relaxation will be indicated
in the following as SH. Indeed, converged calculations should be
yield $b/c=\sqrt{3}$ (SH $=$ sh) after relaxation. For the
{\it Imma} phase, all lattice and internal parameters of the BCO
structure were allowed to relax.

\begin{figure}[ht]
  \epsfig{figure=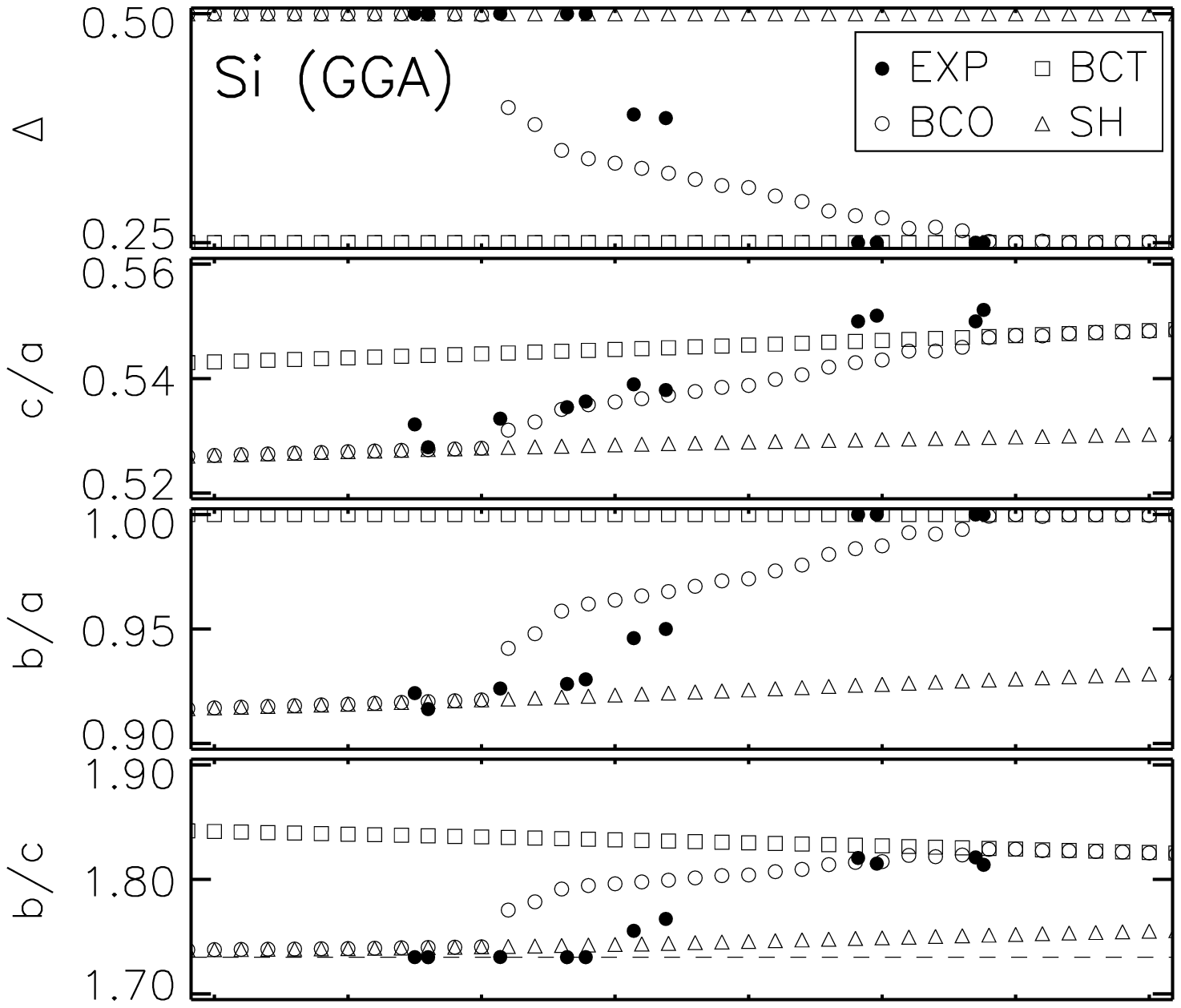, width=8.6cm,angle=0}
  \vspace{0.2cm}
  \epsfig{figure=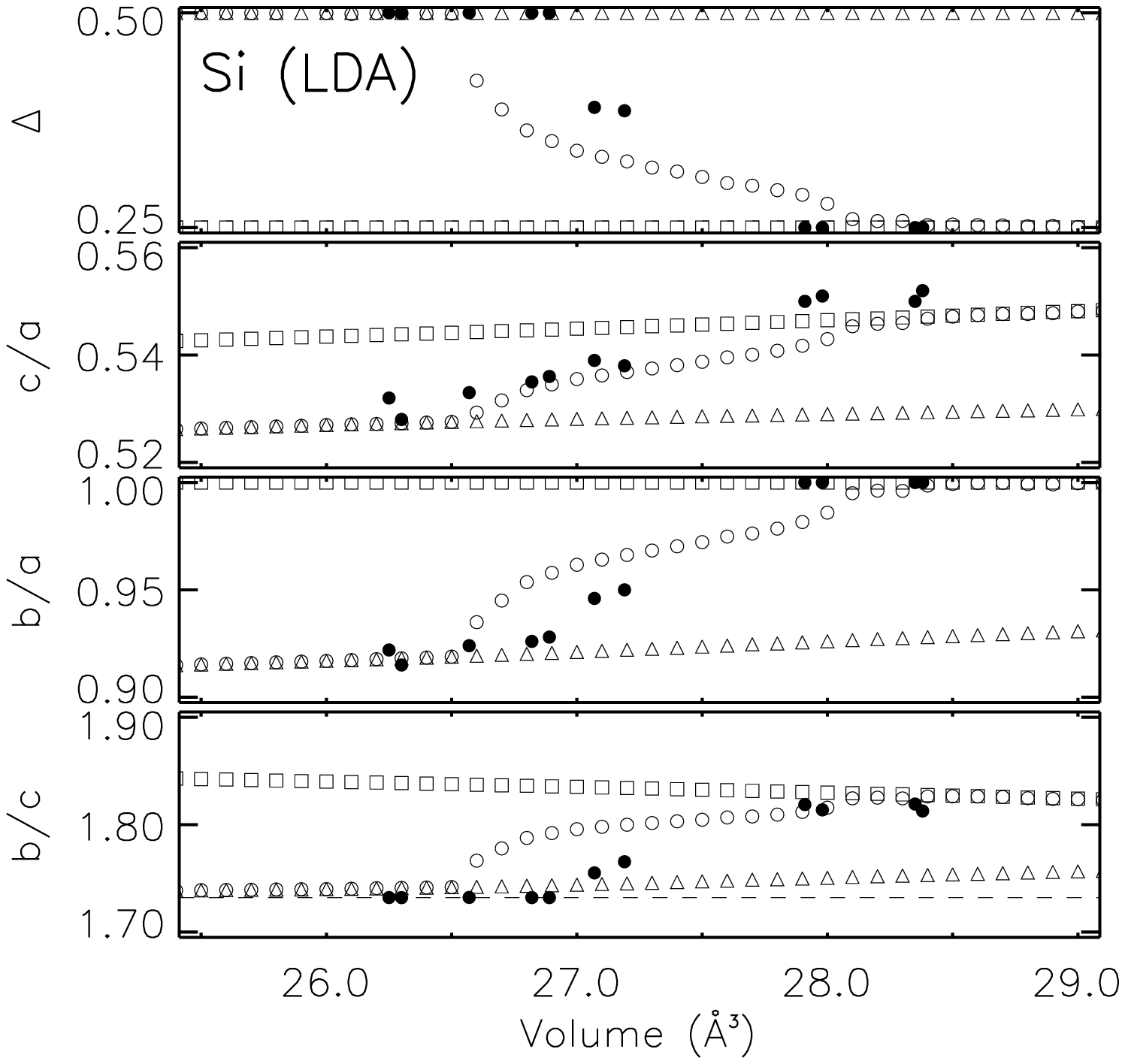, width=8.6cm,angle=0}
  \caption{Volume dependence of the structural parameters of the
  BCT, BCO, and SH structures for Si within GGA (upper
    panels) and LDA (lower panels). The dashed line indicates the
    ideal value of $b/c=\sqrt{3}$. The experimental data are taken
    from
   Refs.~\onlinecite{McMahon1993,McMahon1994,Hu1986,Olijnyk1984,Jamieson1963}.}
  \label{relaxSi}
\end{figure}

\begin{figure}[ht]
  \epsfig{figure=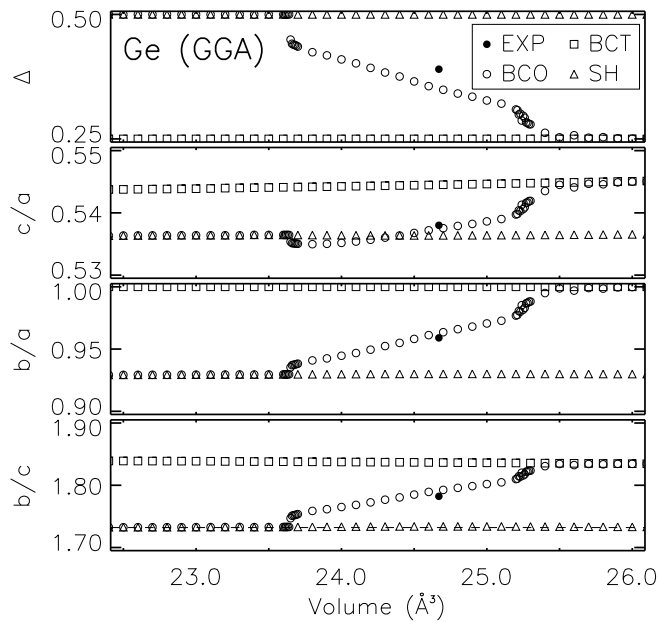, width=8.6cm,angle=0}
  \caption{The same of Fig.~\ref{relaxSi} for Ge within GGA.
    The experimental data are taken from
    Ref.~\onlinecite{Nelmes1996}.}
  \label{relaxGe}
\end{figure}

The relaxed parameters as a function of the volume are displayed in
Figs.~\ref{relaxSi} and \ref{relaxGe} for Si and Ge,
respectively. The results from calculations performed withing the
GGA and the LDA are quite similar for Si. Since it turned out, that
the pressures calculated by GGA are more accurate in Si, only GGA
calculations have been performed for Ge. As can be seen in
Fig.~\ref{relaxSi}, the relaxation of $b/c$ for the SH phase of Si
does not match the ideal value. At variance, the calculation for Ge
does reproduce this value. This indicates that for a full convergence a
higher kinetic-energy cutoff and/or a larger grid of special points
should have been used for Si. However, the condition $b/c=\sqrt{3}$
is almost fulfilled in the region of stability of the sh
phase. Furthermore, the energy obtained from a calculation in the
same cell with $b=\sqrt{3}c$ fixed and $c/a$ relaxed differs from
the value obtained using the previous procedure by less than
0.7~meV. For a more detailed discussion see
Ref.~\onlinecite{Gaal2003}.

The agreement between our data and those experimental ones which are
not included in the figure is good: The reported value of
$c/a=0.552$ for the $\beta$-tin phase of Si\cite{Hu1984} is well
within the range of our values. The experimental values of
$b/a=0.917$ and $c/a=0.530$ at the volume $V=24.43$~\AA$^3$ of
Ref.~\onlinecite{Duclos1990} agree with our data. Furthermore, the
measured value of 24.22~\AA$^3$ for the volume of the sh phase of
Si\cite{Hanfland1999} is also in the stability range we have found
for this phase. For Ge in the BCT structure, a $c/a$ ratio of
0.548--0.554 was detected
experimentally\cite{Jamieson1963,Baublitz1982,Quadri1983,Menoni1986,Vohra1986}
corresponding to a volume region around 33~\AA$^3$; in this region,
our $c/a$ ratio is 0.546--0.549. For the sh phase of Ge, a $b/a$
ratio of 0.93 was measured\cite{Vohra1986} which also agrees with
our data.

Figures \ref{relaxSi} and \ref{relaxGe} show that the relaxation of
the general BCO structure leads to the BCT structure for large
volumes and to the SH structure for small ones. In the intermediate
region, extending from about 26.5 to 28.0~\AA$^3$ for Si and 23.5
from 25.5~\AA$^3$ for Ge, we localise the stability range of the
pure {\it Imma} phase. Whereas the transition between the {\it Imma}
and the sh phase is characterised by a discontinuity in most of the
structural parameters, the {\it Imma} and the $\beta$-tin phase are
quite difficult to distinguish by an inspection of
Figs.~\ref{relaxSi} and \ref{relaxGe}. A clearer picture is obtained
from the electronic structure and the Fermi energy (see
Appendix~\ref{App} for details). From this procedure, the stability
range of the {\it Imma} phase is from 26.6 to 28.3~\AA$^3$ for
Si~(GGA), from 26.6 to 28.0~\AA$^3$ for Si~(LDA), and from 23.7 to
25.4~\AA$^3$ for Ge.

The standard procedure for calculating the volume change and the
pressure at the transition is the so-called common-tangent
construction. It relies on the fact that, in the case of a
first-order phase transition, two structures with the same pressure
$p$ and enthalpy $H$ but different total energy $E(V)$ coexist. The
volume derivative of the total energy is the negative pressure. Therefore,
the slope of the common tangent to the two energy curves gives the
pressure for the transition between the corresponding
structures. Due to the fact that calculations are performed at a
finite number of volumes, a fit of an analytical expression to a
continuous energy curve is 
needed in order to obtain a continuous derivative.

\begin{figure}[ht]
  \epsfig{figure=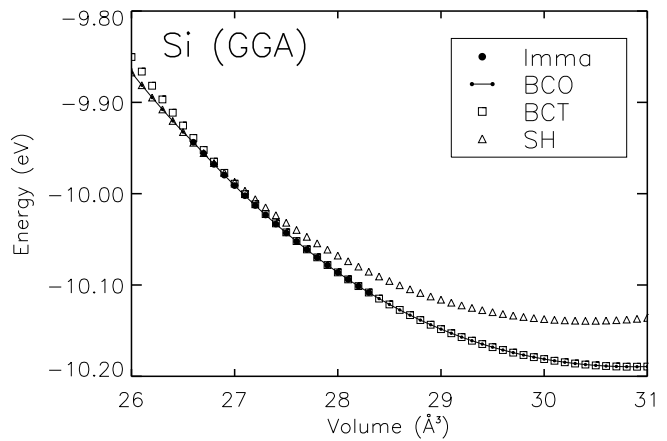, width=8.6cm,angle=0}
  \caption{Energy vs.\ volume for the involved structures of
    Si within GGA (see text for explanation).}
  \label{EV}
\end{figure}

In Fig.~\ref{EV} we show the energy of the relaxed structures for Si
as a function of volume. The volume region where the pure {\it Imma}
phase is stable does not include any minimum of the total energy
$E$. This makes a fit to any equation of state for the {\it Imma}
phase slightly unstable, which results in an unreliable
determination of the transition pressures and of the volume change
through the common-tangent construction in this region. The results
from a fit of the Vinet\cite{Vinet1986} and
Murnaghan\cite{Murnaghan1944} equation of state are shown in
Table~\ref{EOS} for Ge as well as a comparison of the LDA and GGA
calculations for Si.

\begin{table}[h!]
  \caption{Equilibrium parameter determined by the Murnaghan (Mur)
    and Vinet (Vin) equation of state for the cd, $\beta$-tin, {\it
      Imma}, and sh phase of Si and Ge. The equilibrium volume $V_0$
      is in units of \AA$^3$ and the bulk modulus $B_0$ in
      kbar.}\label{EOS}
  \begin{ruledtabular}
    \begin{tabular}{lr|c|c|c|c|c|c}
      & & \multicolumn{2}{c|}{Si-GGA} & \multicolumn{2}{c|}{Si-LDA} & 
      \multicolumn{2}{c}{Ge-GGA} \\
      \hline
      &
      & Mur  & Vin 
      & Mur  & Vin 
      & Mur  & Vin    \\
      \hline
      cd      & $V_0$  
      & 40.65  & 40.61  
      & 39.18  & 39.14  
      & 47.75  & 47.74   \\
      & $B_0$  
      & 892    & 883    
      & 949    & 956    
      & 584    & 593     \\
      & $B'_0$ 
      & 3.67   & 4.21   
      & 3.77   & 4.16   
      & 4.75   & 4.80    \\
      \hline
      $\beta$-tin    & $V_0$  
      & 30.86  & 30.87  
      & 29.59  & 29.58  
      & 38.85  & 38.83   \\
      & $B_0$  
      & 1067   & 1041   
      & 1163   & 1159   
      &  720   & 647     \\
      & $B'_0$ 
      & 3.87   & 4.46   
      & 3.99   & 4.36   
      & 3.77   & 5.10    \\
      \hline
      {\it Imma}     & $V_0$  
      & 31.45  & 31.54  
      & 30.22  & 30.05  
      & 39.03  & 38.66   \\
      & $B_0$  
      & 795  & 792    
      & 799  & 874    
      & 805  & 794       \\
      & $B'_0$ 
      & 4.94 & 4.98   
      & 4.97 & 4.89   
      & 3.31 & 4.39      \\
      \hline
      sh       & $V_0$  
      & 30.42  & 30.43  
      & 29.12  & 29.11  
      & 38.77  & 39.08   \\
      & $B_0$  
      & 1030   & 1012   
      & 1135   & 1137   
      &  768   &  600    \\
      & $B'_0$ 
      & 4.05   & 4.61   
      & 4.20   & 4.52   
      & 3.50   & 5.24    \\
    \end{tabular}
  \end{ruledtabular}
\end{table}

Typically, the GGA tends to overestimate the equilibrium volume
whereas the LDA tends to underestimate it. For the cd phase of Si
the experimental equilibrium volume at room temperature
is\cite{McMahon1993,Bergamin1999,Basile1994,Holloway1991,Windisch1990}
40.05~\AA$^3$, which lies between the LDA and GGA ones, with a
deviation of about 2$\%$. The experimental value for cd Ge
is\cite{Quadri1983,Holloway1991,Yoshiasa1997,Cooper1962} between
45.00 and 45.31~\AA$^3$, with a deviation from our result of less
than 6$\%$. Because the other phases exist only under pressure,
experimental values do not exist to which our equilibrium
parameters could be compared.

If the standard common-tangent construction is used in order to
extract the transition parameters, an accurate determination of
both, equilibrium volumes and curvature of the energy vs.\ volume
curve, is needed. This applies even more so in the present case in
which $E(V)$ curves and equilibrium volumes of the $\beta$-tin, {\it
Imma} and sh phases are very close together. Even though the present
calculation are numerically involved the resulting data of
Table~\ref{EOS} are not precise enough.  For example, in all
calculations (LDA and GGA) for Si the fitted (Vinet and Murnaghan) energy
curves of the {\it Imma} phase fall below the ones of the
$\beta$-tin phase, even for large volumes. In these cases the {\it
Imma} phase is more stable than the $\beta$-tin phase. Thus, within
this construction either a phase transition from the cd to the
$\beta$-tin phase can be observed, if one neglects the {\it Imma}
phase, or a phase transition from the cd to the {\it Imma} phase can
be observed without a locally stable $\beta$-tin phase. A direct
transition from the $\beta$-tin to the {\it Imma} phase cannot be
detected. Similar drawbacks are found for Ge. Furthermore, the
method is applicable only to the investigation of first-order phase
transitions. For all these reasons, it is very difficult to extract
the reliable information on the transitions using the tangent
construction. In the following section, we consider an alternative
procedure for analysing the transitions which does not contain the
drawbacks of the above one.
%
%
\section{Determination of the transition pressures and hysteresis
  effect}\label{pressure} 
In this section we present results for the pressure dependence of
the structural parameters in order to give an estimate of the
transition pressures by considering the stability range of the {\it
Imma} phase.

\begin{figure}[h!]
  \epsfig{figure=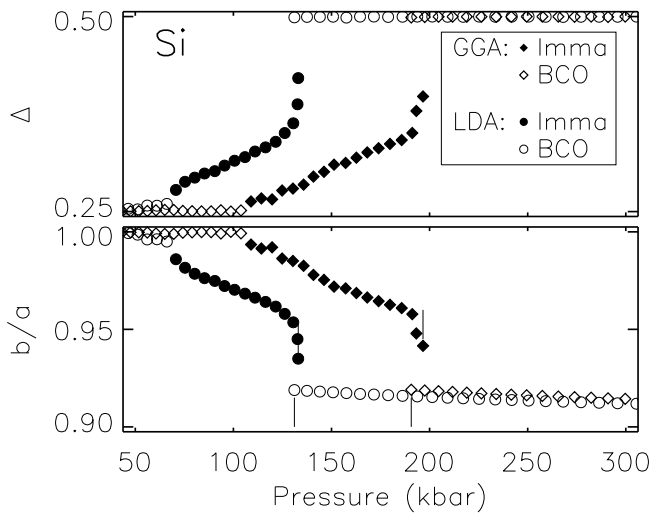, width=8.6cm,angle=0} 
  \caption{Dependence of the structural parameters $\Delta$ and
    $b/a$ on the hydrostatic pressure for the relaxed phases for Si
    with GGA (full and empty diamonds) and LDA (full and empty
    circles). The range of the pressure hysteresis of the
    overlapping phase regions are marked by vertical lines. }
  \label{hysSi}
\end{figure}

\begin{figure}[h!]
  \epsfig{figure=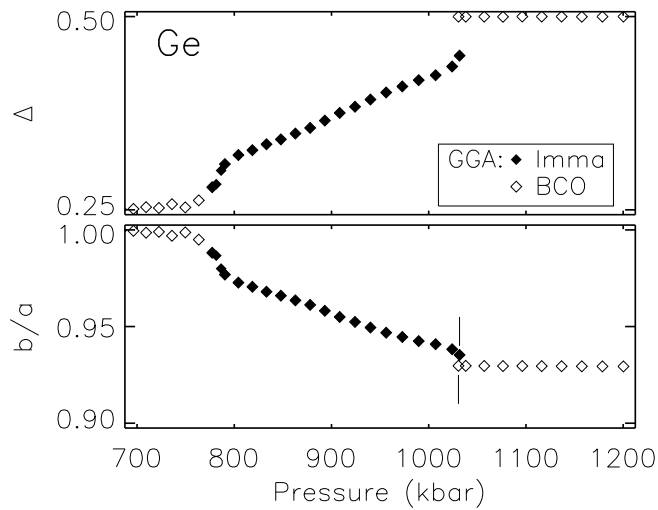, width=8.6cm,angle=0} 
  \caption{The same of Fig.~\ref{hysSi} but for Ge for GGA only.}
  \label{hysGe}
\end{figure}

The pressure $p$ in a crystal can be directly calculated from the
stress tensor {\boldmath $\sigma$}. Within the BCO structure the
off-diagonal components of the stress tensor vanish. Therefore, the
most general form of the stress tensor of our structures is
\begin{eqnarray} \label{stresstensor}
  \mbox{\boldmath $\sigma$} = - \left( 
    \begin{array}{ccc}
      p_x &     &     \\
      & p_y &     \\
      &     & p_z 
    \end{array}
  \right) \quad .
\end{eqnarray} 
The results of the previous section correspond to hydrostatic
pressure
\begin{eqnarray} \label{hydrPress}
  p=p_x=p_y=p_z \quad ,
\end{eqnarray}
and, in our results of Section~\ref{parameters},
Eq.~(\ref{hydrPress}) is fulfilled within an error of 0.1~kbar for
Si and 0.05~kbar for Ge, which is less than the pressure error due
to the Pulay stress. For each of the relaxed BCO structures as
described in Section~\ref{parameters}, we have calculated the
hydrostatic pressure corresponding to the structural parameters as
shown in Figs.~\ref{relaxSi} and \ref{relaxGe}. The pressure
dependence of the structural parameters is presented in
Figs.~\ref{hysSi} and \ref{hysGe}. There is a clear discontinuity of
the parameters of the {\it Imma} phase and the sh phase. The values
of the parameters of the pure {\it Imma} phase are highlighted as
special values of the relaxed BCO structure. For both Si and Ge, the
$\beta$-tin$\to${\it Imma} phase transition seems to be
continuous. For Si, the relaxed parameters calculated within LDA and
GGA are very similar, but the corresponding pressures differ as
shown in Fig.~\ref{hysSi}. In contrast, the volume range of
stability in both cases is similar (see
Fig.~\ref{relaxSi}). Usually, the GGA transition pressures are
closer to the experimental ones than the LDA values. Therefore, we
will focus in the following mainly on the GGA results.


The assignment of the phases from the data points in
Figs.~\ref{hysSi} and \ref{hysGe} are taken from
Section~\ref{parameters}.  In the case of the GGA calculation for Si
we have found the following sequence with pressure increase: For the
$\beta$-tin$\to${\it Imma} phase transition, the transition pressure
is approximately between 104 (last relaxation of the system to a BCT
structure) and 109~kbar (first relaxation of the system to a BCO
structure with $\Delta\neq0.25$). The next transition is the one
from the {\it Imma} to the sh phase. As can be seen in
Figs.~\ref{hysSi} and \ref{hysGe}, there is an overlap of different
structures for certain values of the pressure, which can be
interpreted as a hysteresis effect. That means that the transition
value for increasing and decreasing pressure is different. This
effect was first observed first in the experiment for
Si.\cite{McMahon1994} We estimate the hysteresis effect for Si as
follows: The last appearance of the pure {\it Imma} phase
($\Delta\neq0.5$) {\sl without} any SH structure at the same
pressure is at 186~kbar, the first appearance of the sh phase is at
191~kbar (downstroke), the last appearance of the pure {\it Imma} phase is at
197~kbar and the first appearance of the sh structure {\sl without}
any parallel {\it Imma} phase is at 198~kbar (upstroke). The transition
pressures estimated in this way are listed in Table~\ref{trans}.

\begin{table}[ht!]
  \caption{Summary of the estimated transition pressures at
    increasing ($\uparrow$) and decreasing ($\downarrow$) pressure
    for the phase transitions between the $\beta$-tin, {\it Imma}
    and sh phase.}\label{trans}
  \begin{ruledtabular}
    \begin{tabular}{cccr|cc|cc}
      \multicolumn{4}{l}{Si}&
      \multicolumn{4}{c}{transition pressure (kbar) } \\ 
      \hline 
      $\beta$-tin & $\leftrightarrow$ & {\it Imma} & GGA &
      \multicolumn{2}{c}{104--109} & 
      \multicolumn{2}{c}{($\uparrow\downarrow$)} \\ 
      & & & LDA &
      \multicolumn{2}{c}{66--71} & 
      \multicolumn{2}{c}{($\uparrow\downarrow$)} \\ 
      \hline
      {\it Imma} & $\leftrightarrow$ & sh & GGA & 
      197--198 & ($\uparrow$) & 186--191 & ($\downarrow$) \\ 
      & & & LDA & 
      133--138 & ($\uparrow$) & 131 & ($\downarrow$) \\ 
      \hline
      \hline
      \multicolumn{4}{l}{Ge}&
      \multicolumn{4}{c}{transition pressure (kbar) } \\
      \hline 
      $\beta$-tin & $\leftrightarrow$ & {\it Imma} & GGA &
      \multicolumn{2}{c}{750--763} &
      \multicolumn{2}{c}{($\uparrow\downarrow$)} \\ 
      \hline
      {\it Imma} & $\leftrightarrow$ & sh & GGA & 
      1032 & ($\uparrow$) & 1031--1029 & ($\downarrow$) \\ 
    \end{tabular}
  \end{ruledtabular}
\end{table}

Since we have calculated the lowest energy structure, the values in
Table~\ref{trans} give an estimate of the hysteresis effect. An
alternative approach to the hysteresis effect will be described in
Section~\ref{barriers}.
%
%
\section{Order of the phase transitions}\label{order}
The thermodynamical potential for a pressure-induced phase
transition is the Gibbs free energy. Neglecting temperature effects,
the Gibbs free energy is equal to the enthalpy $H=E+pV$. For
calculating the enthalpy, we have used the energy $E(V)$ of the
relaxed structures with a volume $V$ and the hydrostatic pressure
$p$ from Eq.~(\ref{hydrPress}).

At a given pressure, the stable structure is the one with the lowest
enthalpy. Therefore, the transition pressure $p_t$ for a transition
$A${$\to$}$B$ is defined by $H^A(p_t)=H^B(p_t)$. For a first-order
phase transition the enthalpy curves $H^A(p)$ and $H^B(p)$ of the
two phases cross at $p_t$, and their derivatives ${\rm d} H/{\rm d}
p$ at $p_t$ are different for the two phases.  For a second-order
transition, the second derivatives of the enthalpy curves at $p_t$
for the two phases are different, whereas the first derivatives are
equal. In this case the corresponding enthalpy curves have a
boundary point at $p_t$.

\begin{figure}[ht]
  \epsfig{figure=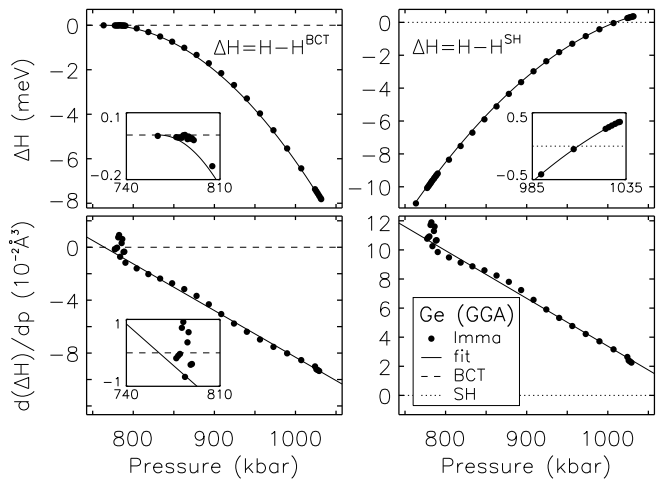, width=8.6cm,angle=0} 
  \caption{Enthalpy vs.\ pressure (upper panels) and derivative of
    the enthalpy vs.\ pressure (lower panels) for Ge reduced
    by the corresponding curves of the BCT (right panels) and the SH
    (left panels) structure. The solid lines are quadratic (upper
    panels) and linear (lower panels) fits.} \label{GeOrder}
\end{figure}

In the following, we analyse the transitions in order to extract the
corresponding transition pressure and order. The enthalpy $H(p)$ and
its first derivative are presented in Fig.~\ref{GeOrder} for Ge.
The enthalpy curves have been reduced by the value of the BCT
structure for the $\beta$-tin$\to${\it Imma} transition,
\begin{eqnarray}
\Delta H(p) = H(p) - H^{\rm BCT}(p)
\end{eqnarray}
and by the value of the SH structure for the {\it Imma}$\to$sh transition,
\begin{eqnarray}
\Delta H(p) = H(p) - H^{\rm SH}(p) \quad .
\end{eqnarray}
The derivative indicated by black dots in the lower panels of
Fig.~\ref{GeOrder} has been calculated by finite differences, the
straight solid line is a linear fit of these data, obtained by using
only points belonging to the numerically stable regime
(790--1030~kbar). An integration of this linear function is fully
consistent with the enthalpy values in the upper panels. The
enthalpy differences of the points in the pressure range
$p<790$~kbar stem from the numerical noise of our calculation.

For the {\it Imma}$\to$sh transition (right panels of
Fig.~\ref{GeOrder}) the two enthalpy curves cross at the transition
pressure $p_t=1009$~kbar. Therefore, this phase transition is
clearly of first order. The change of the derivatives at the
transition point is the volume change $\Delta V(p_t)$ and can be
directly read from the lower panel of Fig.~\ref{GeOrder}.

For $\beta$-tin$\to${\it Imma} (left panels of Fig.~\ref{GeOrder})
the enthalpy curves are very close to each other in the region where
the phase transition is expected. From this point of view, transitions
of both first and second order are conceivable. We have thus
analysed our data assuming alternatively a first- and a second-order
transition.

Assuming a second-order transition the transition pressure is
obtained by looking at the crossing of the ${\rm d} H/{\rm d} p$
curves.

A precise value of the $\beta$-tin$\to${\it Imma} transition
pressure assuming first order can not be determined by a crossing of
the enthalpy curves due to the numerical accuracy of our
calculation. We are only able to give a rough estimate of the
pressure at which the transition can occur due to an extrapolation
of the more reliable data points away from the transition
pressure. The results for Si have been obtained in an analogous way.

For both Si and Ge, the values of $p_t$ and of the volume changes
between the low- and the high-pressure phase are shown in
Table~\ref{transorder}.

\begin{table}[ht]
  \caption{Transition pressure ($p_t$), order (ord), and volume change
    ($\Delta V$) for Ge from Fig.~\ref{GeOrder} and similarly for
    Si.}\label{transorder}
  \begin{ruledtabular}
    \begin{tabular}{ll|c|cc|cc}
      & & & \multicolumn{2}{c|}{$p_t$ (kbar)} & 
      \multicolumn{2}{c}{$\Delta V$(m\AA$^3$)} \\
      & & ord & GGA & LDA & GGA & LDA  \\
      \hline
      cd$\to${$\beta$}-tin 
      &Si& 1   & 121      & 79 & 8136 & 8583 \\ 
      &Ge& 1   & 96       & -- & 7508 &  \\ 
      \hline
      $\beta$-tin$\to${\it Imma} 
      &Si& 2   & 108      & 65  & 0     & 0    \\ 
      &Si& 1   & 103      & 71  & $-1.2$  & 10.1 \\ 
      &Ge& 2   & 765      & --  & 0     & --   \\ 
      &Ge& 1   & 792      & --  & 9.5   & --   \\ 
      \hline
      {\it Imma}$\to$sh 
      &Si& 1   & 189      & 127 & 301.4 & 317.3 \\ 
      &Ge& 1   & 1009     &  -- &  30.7 & --    \\ 
    \end{tabular}
  \end{ruledtabular}
\end{table}

For Si the volume changes $\Delta V$ for an assumed first-order
$\beta$-tin$\to${\it Imma} transition from LDA and from GGA and are
very small compared to 
the changes for {\it Imma}$\to$sh. Thus we conclude that the phase
transition $\beta$-tin$\to${\it Imma} in Si most probably is of
second order.

For Ge the volume changes for a first-order $\beta$-tin$\to${\it
Imma} and {\it Imma}$\to$sh transition are within the same
range. The value of $p_t=792$~kbar is higher than the ones from
the estimated range (750--763~kbar) of the previous section, and
overestimating $p_t$ results in overestimating $\Delta V$. A reason
for the large value of $p_t$ can be found in the fact, that the last
reliable data points are too far away from the region where the
phase transition is expected to occur. Thus, it is not possible to
decide whether the transition $\beta$-tin$\to${\it Imma} is of first
or of second order. Because of the similarities of Si and Ge a
second-order transition seems more probable. 
If this phase
transition is a discontinuous one, it is at least weakly of first
order. 

Using the enthalpy method the transition pressure for
cd$\to${$\beta$}-tin 
in Si is calculated to be larger than the one for the
$\beta$-tin$\to${\it Imma} transition. These two phase transitions
take place in a small pressure range and therefore numerical errors
in the determination of the transition pressure may lead to a wrong
sequence. See the discussion at the end of Section~\ref{parameters}.
%
%
\section{Energy surfaces and enthalpy barriers}\label{barriers}
In this section, we present the results for the energy surfaces for
Ge (similar calculations have been done for Si) as well as for the
enthalpy barriers for the cd$\to${$\beta$}-tin and {\it Imma}$\to$sh
transitions. 

We have obtained the energy surface by calculating the energy
corresponding to given values of the parameters $V$, $b/a$, and
$c/a$ and a relaxed internal parameter $\Delta$. On the energy
surface the pressure is non-hydrostatic except one or two lines or a
few points. Next, we
consider the pressure defined in Eq.~(\ref{hydrPress}) in the
non-hydrostatic case
\begin{eqnarray}\label{averageP}
p_0=\frac{1}{3} \left(p_x+p_y+p_z\right)
\end{eqnarray}
where $-p_x$, $-p_y$, and $-p_z$ are the components of the diagonal
stress tensor, Eq.(\ref{stresstensor}).

For every first-order phase transition, there exists an energy
barrier which has to be overcome on the path from the one phase to
the other. For most reactions, such as, e.g., adsorption processes,
it is a barrier of the total energy $E$ which lies between two local
minima in the corresponding energy space. For this kind of
transition, the reaction path and the energy barrier between the two
phases can be found by using, e.g., the nudged-elastic-band
method,\cite{Mills1995,Jonsson1998} by which the lowest energy path
between two local minima is detected. However, this method requires
the existence of well-defined local energy minima corresponding to
the two phases.

\begin{figure}[ht]
  \epsfig{figure=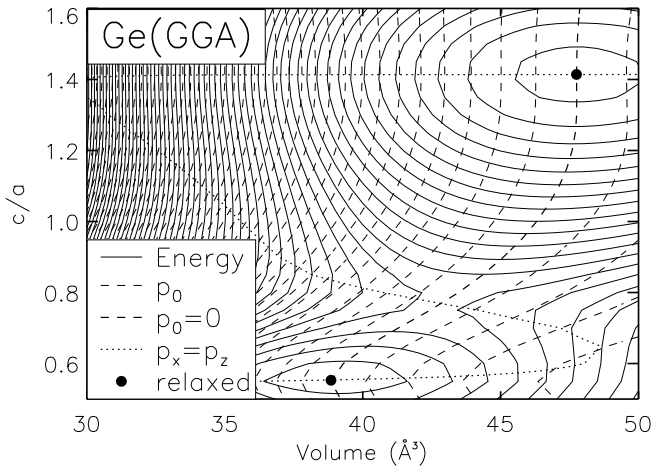,width=8.6cm,angle=0}
  \caption{Contour plot of the total energy $E(V, c/a)$ and of the
    average pressure $p_0(V,c/a)$ (see Eq.~\ref{averageP}) for Ge. The
    interval of the contour lines is 50~meV for the energy and
    20~kbar for the pressure surfaces. The conditions $p_x=p_y$,
    $b/a=1$, and $\Delta=0.25$ are fulfilled within the whole
    area. The black dots mark the equilibrium positions of the cd
    ($c/a=\sqrt{2}$) and the $\beta$-tin phase ($c/a=0.55$). The
    dotted line markes the parameters under hydrostatic
    condition.}\label{GefccE} 
\end{figure}

If the transition, for instance the one from the cd to the
$\beta$-tin phase in Ge, would occur without any influence of
pressure, the corresponding energy space would be the energy surface
$E(V,c/a)$ drawn in Fig.~\ref{GefccE}. The reaction coordinate here
is taken as $c/a$. As expected, there are two local minima and a
saddle point between them. By symmetry, the condition $p_x=p_y$ is
valid all over 
the energy surface. The energy minima and the saddle point lie on
the contour line $p_x=p_z$, which indicates the hydrostatic
condition. The energy barrier is then defined as the energy
difference between the saddle point and the starting minimum.

\begin{figure}[ht]
  \epsfig{figure=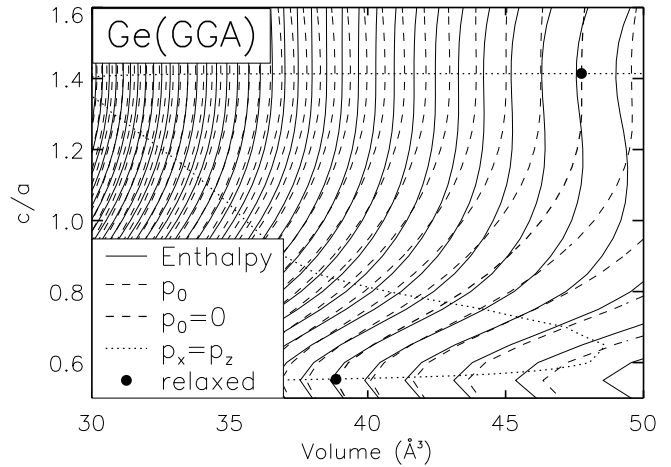,width=8.6cm,angle=0}
  \caption{Same as Fig.~\ref{GefccE} but for the enthalpy $H(V,
    c/a)$ of Ge. The interval between the contour lines for $H$ is
    500~meV.}
  \label{GefccH}
\end{figure}

However, in our case the phase transition is driven by the pressure,
and therefore the associated thermodynamical potential is the Gibbs
free energy or the enthalpy, if temperature effects are
neglected. The enthalpy surface corresponding to the cd and
{$\beta$}-tin phases for Ge is shown in Fig.~\ref{GefccH}. The
enthalpy for each set of parameters ($V,c/a$) is calculated as
$H(V,c/a)=E(V,c/a)+V\,p_0(V,c/a)$. In the range displayed in
Fig.~\ref{GefccH} the enthalpy surface has no minimum. Therefore,
the usual path methods are not applicable here. However, the
enthalpy barrier can be calculated in a different way as follows:
The phase transition between the $A$ and $B$ phases occurs at the
pressure $p_t$ at which $H^A(p_t)=H^B(p_t)$. Hence, we have to look
for the points along the isobars in Fig.~\ref{GefccH} which have the
same value of the enthalpy.

\begin{figure}[hb!]
  \epsfig{figure=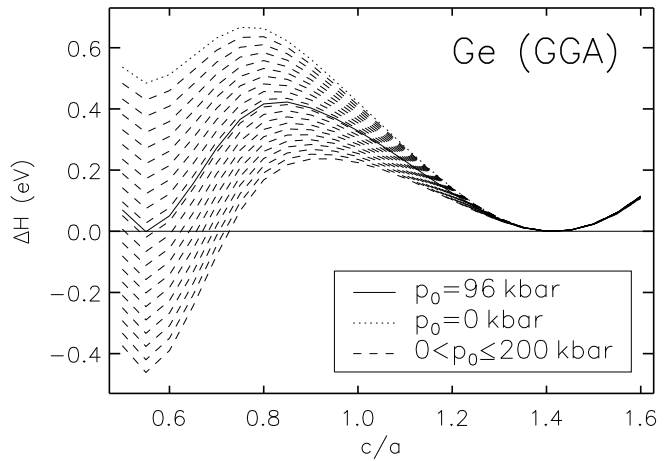,width=8.6cm,angle=0}
  \caption{Enthalpy difference at constant average pressure vs.\ the
    reaction coordinate $c/a$, see Eq.~(\ref{normfcc}) for Ge. The solid
    line is for the cd$\to${$\beta$}-tin transition pressure. The
    pressure interval between consecutive lines is 10~kbar.}
  \label{GefccBarr} 
\end{figure}

In order to visualise the enthalpy barrier, we draw in
Fig.~\ref{GefccBarr} the enthalpy at constant average pressure (with
$p_x=p_y$) as a function of the reaction coordinate $c/a$. For
convenience the enthalpy is reduced by the starting point with
$c/a=\sqrt{2}$,
\begin{eqnarray} \label{normfcc}
\Delta H (c/a) = H(c/a)-H(\sqrt{2}) \quad ,
\end{eqnarray}
which determines here the cd structure. As can be seen in
Fig.~\ref{GefccBarr}, there exists just one enthalpy line at
constant pressure, which touches the zero line twice according to
$H^A(p)=H^B(p)$. The corresponding pressure is the transition
pressure $p_t$ and the height of the maximum between these minima is
the height of the enthalpy barrier for this phase transition. Note
that on the isobars at most three points refer to hydrostatic pressure,
while the other points corresponds just to the condition
$p_x=p_y$. 

Similar pictures can be drawn for the
cd$\to${$\beta$}-tin transition of silicon. The height of enthalpy
barrier is found to be 423~meV for Ge within GGA and 515 and 508~meV
for Si within GGA and LDA, respectively. In agreement with previous
results,\cite{Gaal2002} the barrier for Ge is smaller than for Si.

\begin{figure}[ht]
  \epsfig{figure=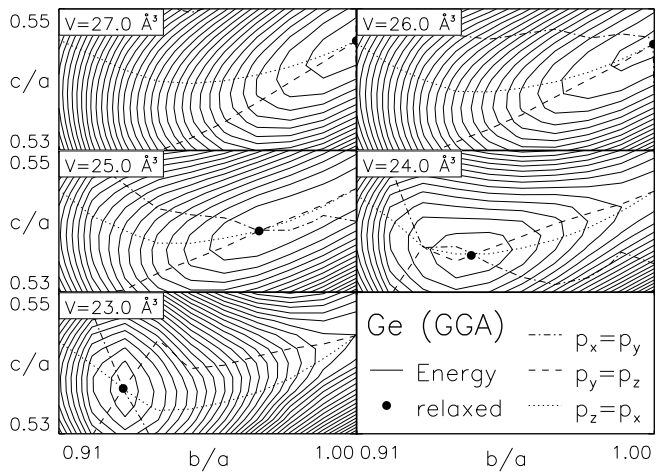, width=8.6cm,angle=0} 
  \caption{Contour plots of the total energy $E(b/a, c/a)$ for
    selected volumes noted in the insets for Ge. The interval of the
    contour lines is 1~meV. The black dots mark the equilibrium
    positions. The hydrostatic condition is produced by the crossing
    of all the lines $p_x=p_y$, $p_y=p_z$, and $p_z=p_x$.}
  \label{GebcoE}
\end{figure}

For the cd{$\to$}{$\beta$}-tin phase transition the determination of
the enthalpy barrier is easy because only the volume and the $c/a$
ratio can vary, and the latter can be used as the reaction
coordinate. For the $\beta$-tin$\to${\it Imma}$\to$sh transition all
the parameters $c/a$, $b/a$, $b/c$, and $\Delta$ are changing (but
only two of the ratios are independent). We have calculated the
total energy for each set of parameter ($V,c/a,b/a$), the internal
parameter $\Delta$ has been relaxed for each set. Selected energy
contour plots at constant volume are presented in
Fig.~\ref{GebcoE}. In these plots, the places where two of the
diagonal components of the stress tensor are equal are explicitly
shown by the dotted, dashed, and dash-dotted lines. Obviously,
hydrostatic condition is indicated by the points where all of these
lines cross. Different hydrostatic pressures at one volume (like at
$V=24$~\AA$^3$) can lead to a multivalued pressure-volume relation.

\begin{figure}[ht]
  \epsfig{figure=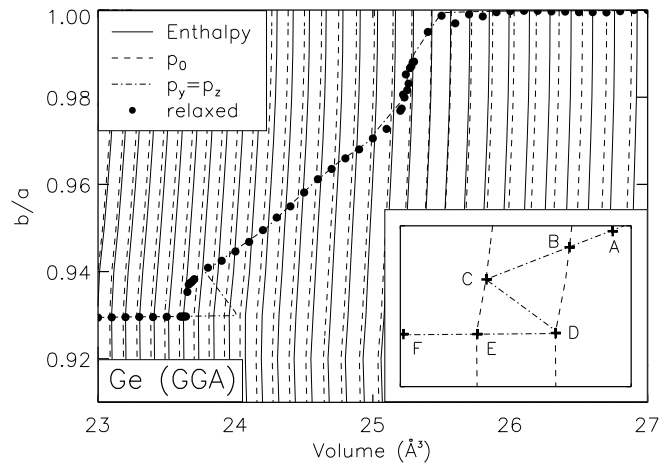, width=8.6cm,angle=0} 
  \caption{Contour plot of the enthalpy $H(V, b/a)$ and of the
    pressure $p_0(V,b/a)$ for Ge. The interval of the contour lines is
    300~meV for the enthalpy and 20~kbar for the pressure increasing
    from right to left. The condition $p_x=p_z$ is fulfilled within
    the whole area. The dash-dotted line reproduce the hydrostatic
    condition. The inset shows the schematical behavior between 23.5
    and 24~\AA$^3$, see text.} \label{GebcoH}
\end{figure}

In order to calculate the enthalpy barrier in analogy to the
cd$\to${$\beta$}-tin transition, we present a contour plot of the
enthalpy as a function of the volume $V$ and of the $b/a$ ratio in
Fig.~\ref{GebcoH}. Actually, the enthalpy depends also on the other
variables $c/a$ and $\Delta$. The contour plot of Fig.~\ref{GebcoH}
has been obtained using the following procedure: For each value of
the set ($V,b/a$), we have chosen $c/a$ so as to fulfill the
condition $p_z=p_x$ (see Fig.~\ref{GebcoE}), which is suggested
because at the end we will look for the points which fulfill the
hydrostatic condition. The value of $\Delta$ has been taken so as to
minimise the energy for a given choice of the other parameters. The
two-dimensional presentation of the results as in Fig.~\ref{GebcoH}
corresponds to the choice of $b/a$ as the reaction
coordinate. Completely equivalent results can be obtained by
choosing $c/a$ or $\Delta$ instead. The hydrostatic condition in
Fig.~\ref{GebcoH} is then realised by $p_y=p_z$.

The comparison with our previous results shows that most of the data
shown in Fig.~\ref{hysGe} lie on the hydrostatic line. Numerical
instabilities lead to deviations (generally of less than 1~kbar)
between the relaxed points and the line $p_y=p_z$. Nevertheless, the
relaxed points show the same behaviour (to a lesser extent) as
the hydrostatic line, which is shown schematically in the inset of
Fig.~\ref{GebcoH}.

The schematical behaviour of the hydrostatic curve between
approximately 23.5 and 24~\AA$^3$ as shown in the inset of
Fig.~\ref{GebcoH} can be used to discuss the hysteresis effect
qualitatively. If one follows the hydrostatic line $p_y=p_z$ with
increasing pressure $p$ (decreasing volume) starting from A, one
arrives at the point C where the direction of the line is
changing. From C to D, the pressure decreases with a simultanous
decrease of the volume which is an instable situation. Thus, at C
the phase transition must occur through a jump from C to E, which is
at the same pressure. From E increasing pressure leads to
F. Following the path in the opposite direction starting from F,
i.e., by decreasing the pressure (volume increase), the direction of
the curve changes at D, which results in a jump to B. As a
consequence of that it is possible to have a higher transition
pressure with increasing than with decreasing pressure. Therefore,
an alternative estimate for the limit of the hysteresis effect can
be given. For Si within GGA (LDA), we obtaine a transition pressure
of 198~(133)~kbar with increasing and of 169~(109)~kbar with
decreasing pressure. For Ge within GGA, the corresponding values are
1010 and 965~kbar.

\begin{figure}[ht]
  \epsfig{figure=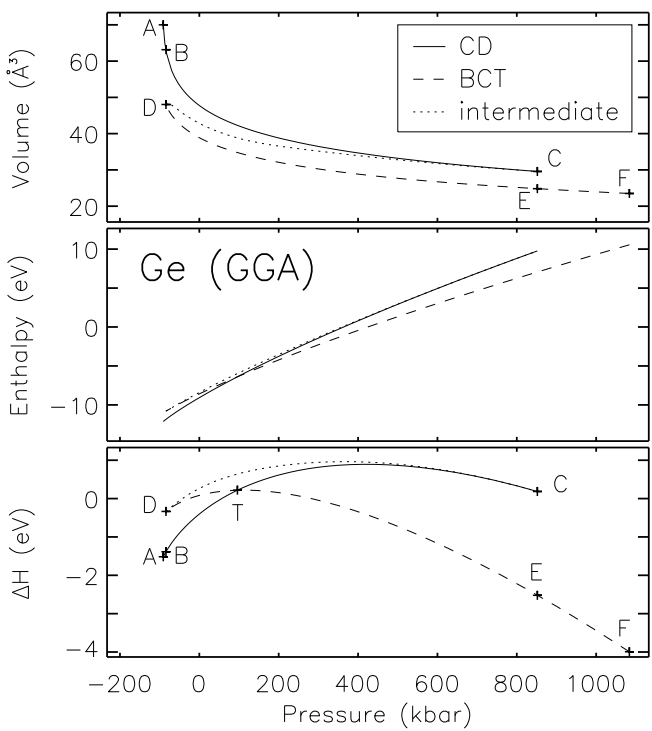, width=8.6cm,angle=0} 
  \caption{Volume, enthalpy and reduced enthalpy vs.\ hydrostatic
    pressure for Ge. The points noted in the figures correspond to
    the points of the inset of Fig.~\ref{GebcoH}; T marks the point
    of the phase transition between the cd and the $\beta$-tin
    phase, see text.}
  \label{GePVHP}
\end{figure}

A behaviour similar to the one depicted in the inset of
Fig.~\ref{GebcoH} can be found also for the cd$\to${$\beta$}-tin
phase transition (see Fig.~\ref{GefccH}). There, the low-pressure
edge corresponding to the point C of the inset of Fig.~\ref{GebcoH} is at a
negative pressure, which would lead for $\beta$-tin$\to$cd to a very
low or even negative transition pressure, the latter case being
connected to the irreversibility of the phase transition. For the
cd$\to${$\beta$}-tin transition the volume and the enthalpy as a
function of hydrostatic pressure is presented in
Fig.~\ref{GePVHP}. The values have been extracted from
Fig.~\ref{GefccH} along the hydrostatic line $p_x=p_z (=p_y)$. The
ideal cd structure ($c/a=\sqrt{2}$) has been reached within an error
of 1\% and therefore cd is noted as CD in analogy to the difference
between sh and SH (see Section~\ref{parameters}). In order to
discriminate the enthalpy curves of Fig.~\ref{GePVHP} we have
subtracted a linear background to arrive at the curves $\Delta H$.
The solid and the dashed line mark the CD and the BCT structure,
respectively. The dotted line indicate the values along the part of
the 
hydrostatic line, which connects the CD and the BCT structure. The
designated points correspond to the inset of Fig.~\ref{GebcoH}.

The condition of local stability is fulfilled along all curves shown
in Fig.~\ref{GePVHP}. Thus, the volume vs.\ pressure curves are
strictly monotonically decreasing and the enthalpy vs.\ pressure
curves are convex. The condition of global stability requires the
enthalpy to be minimal, and so the curves from A to T and from T to F
mark the globally stable regime. The point of the phase transition
in analogy to the Maxwell construction is noted as T. The pressure
range of the coexistence regime is determined by the overlap of the
globally stable and the instable regions, which is here from B (or
D) to E (or C).

A similar picture could be drawn for the {\it Imma}$\to$sh
transition, but with less resolution. A reliable identification of
the structures within the $H(p)$ curve analoguos to
Fig.~\ref{GePVHP} requires a finer grid of data points for the
interpolation and/or a higher convergence.

\begin{figure}[ht]
  \epsfig{figure=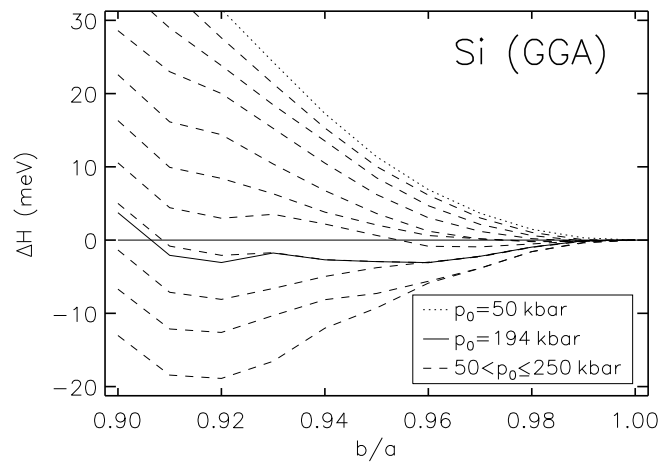, width=8.6cm,angle=0}
  \caption{Enthalpy difference at constant average pressure vs.\ 
    reaction coordinate $b/a$, see Eq.~(\ref{normbct}) for Si. The
    pressure interval between the dashed lines is 20~kbar. The solid
    line marks the transition pressure found here.}
  \label{SibcoBarr}
\end{figure}

\begin{figure}[ht]
  \epsfig{figure=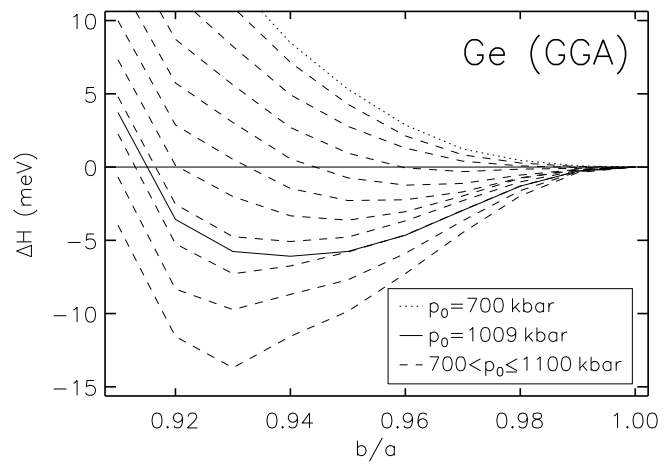, width=8.6cm,angle=0} 
  \caption{Same as Fig.~\ref{SibcoBarr} but for Ge. The pressure
    interval between the dashed lines is 40~kbar. The solid curve
    marks the line corresponding to the transition pressure for the
    {\it Imma}$\to$sh transition obtained in Section~\ref{order}.}
  \label{GebcoBarr}
\end{figure}

The enthalpy barrier for the cd$\to${$\beta$}-tin phase transition
can be determined with a procedure like illustrated in
Fig.~\ref{GefccBarr}. In Figs.~\ref{SibcoBarr} and \ref{GebcoBarr},
we show the enthalpy at given pressures calculated within GGA as a
function of the reaction coordinate $b/a$, for Si and Ge,
respectively.


The enthalpy in Figs.~\ref{SibcoBarr} and \ref{GebcoBarr} is reduced
by the value at $b/a=1$ (corresponding to the $\beta$-tin phase),
\begin{eqnarray}\label{normbct}
\Delta H (b/a) = H(b/a)-H(1) \quad .
\end{eqnarray}
The dotted line in Fig.~\ref{SibcoBarr} corresponds to a pressure
where the enthalpy has a local minimum for the $\beta$-tin phase. At
a pressure of about 130~kbar we find a continuous transition to the
{\it Imma} phase. Above this pressure, a minimum appears at
$b/a<1$. By further increasing $p_0$ a second minimum apprears at
$b/a\sim 0.92$, corresponding to the sh phase. At the pressure of
194~kbar the two minima have the same enthalpy, therefore a
transition can occur. The enthalpy barrier between the two phases is
surprisingly small, namely 1.2~meV. At higher pressure the sh phase
is stable. For the LDA calculation a transition pressure of 133~kbar
and an enthalpy barrier of 1.3~meV are obtained.  If one considers
the enthalpy curves of Fig.~\ref{SibcoBarr} at the transition
pressures determined in the Sections~\ref{pressure} and \ref{order},
one can give a further estimate of the enthalpy barriers, and, the
values obtained following this procedure do not differ appreciably
from the ones mentioned above.

As can be seen in Fig.~\ref{GebcoBarr} for Ge, a minimum of the
enthalpy can be identified corresponding to a pure {\it Imma} phase,
slightly below 820~kbar. However, within our resolution (about
0.1~meV) the pressure curves have only one minimum. Therefore, an
enthalpy barrier between the {\it Imma} and the sh phase can not be
identified. Indeed, this transition is of first order. Hence, the
existing barrier must be smaller than 0.1~meV.

All the enthalpy barriers which we are able to obtain for the {\it
Imma}$\to$sh transition are much smaller than the thermal energy at
room temperature (25~meV). Thus, experiments performed at room
temperature should not be able to distinguish between a first- or a
second-order phase transition.
%
%
\section{Discussion of the results for the phase
  transitions}\label{discussion} 
The transition pressures for the cd$\to${$\beta$}-tin$\to${\it
Imma}$\to$sh phase transitions have been calculated with different
methods in the previous sections. In Sections~\ref{pressure} and
\ref{barriers} an estimate of the hysteresis range was given by
considering the relaxed values (R) and the hydrostatic line (HL)
respectively. In Section~\ref{order} the transition pressure was
determined via the enthalpy vs.\ pressure curves (EC). In
Table~\ref{summary} an overview of all our results for the
transition pressure is given in comparison with the experimental
results.

\begin{table}[ht]
  \caption{Summary of the estimated transition pressures ($p_t$) between
    the cd, $\beta$-tin, {\it Imma}, and sh phase for increasing
    ($\uparrow$) and decreasing ($\downarrow$) pressure, determined
    by the enthalpy curves (EC), the relaxation (R), and the
    hydrostatic line (HL).
    }\label{summary}
  \begin{ruledtabular}
    \begin{tabular}{llc|cccc}
      Si & Method & &
      \multicolumn{4}{c}{$p_t$ (kbar) } 
      \\ 
      \hline 
      cd$\to${$\beta$}-tin 
      & EC & GGA 
      & \multicolumn{1}{c}{ } 
      & \multicolumn{2}{c}{121} & 
      \\ 
      & EC & LDA 
      & \multicolumn{1}{c}{ } 
      & \multicolumn{2}{c}{79}  & 
      \\ 
      \hline
      && exp \footnotemark[1]
      & \multicolumn{1}{c}{($\uparrow$)} 
      & \multicolumn{2}{c}{103-133} & 
      \\
      \hline
      $\beta$-tin$\leftrightarrow${\it Imma}
      & R  & GGA 
      & \multicolumn{1}{c}{($\uparrow\downarrow$)} 
      & \multicolumn{2}{c}{104--109} & 
      \\ 
      & EC & GGA 
      & \multicolumn{1}{c}{ } 
      & \multicolumn{2}{c}{108} & 
      \\ 
      & R  & LDA 
      & \multicolumn{1}{c}{($\uparrow\downarrow$)} 
      & \multicolumn{2}{c}{66--71} & 
      \\ 
      & EC & LDA 
      & \multicolumn{1}{c}{ } 
      & \multicolumn{2}{c}{71} & 
      \\ 
      \hline
      && exp \footnotemark[2]
      & ($\uparrow$) 
      & 134--148
      & 127--131 
      & ($\downarrow$) 
      \\
      \hline 
      {\it Imma}$\leftrightarrow$sh 
      & R  & GGA 
      &($\uparrow$) 
      & 197--198 
      & 186--191 
      & ($\downarrow$) 
      \\ 
      & HL & GGA 
      & ($\uparrow$) 
      & 198 
      & 169 
      & ($\downarrow$) 
      \\ 
      & EC & GGA 
      & \multicolumn{1}{c}{ } 
      & \multicolumn{2}{c}{189} & 
      \\ 
      & R  & LDA 
      & ($\uparrow$) 
      & 133--138 
      & 131 
      & ($\downarrow$) 
      \\ 
      & HL & LDA 
      & ($\uparrow$) 
      & 133 
      & 109 
      & ($\downarrow$) 
      \\ 
      & EC & LDA 
      & \multicolumn{1}{c}{ } 
      & \multicolumn{2}{c}{127} & 
      \\ 
      \hline
      &&exp \footnotemark[2]
      & ($\uparrow$) 
      & 149--154 
      & 140--157 
      & ($\downarrow$) 
      \\
      \hline
      \hline
      Ge & Method & &  
      \multicolumn{4}{c}{$p_t$ (kbar) } 
      \\
      \hline 
      cd$\to${$\beta$}-tin 
      & EC & GGA 
      & \multicolumn{1}{c}{ } 
      & \multicolumn{2}{c}{96} & 
      \\ 
      \hline
      && exp \footnotemark[3]
      & \multicolumn{1}{c}{($\uparrow$)} 
      & \multicolumn{2}{c}{103--110} & 
      \\
      \hline 
      $\beta$-tin$\leftrightarrow${\it Imma} 
      & R & GGA 
      & \multicolumn{1}{c}{($\uparrow\downarrow$)} 
      & \multicolumn{2}{c}{750--763}& 
      \\ 
      & EC & GGA 
      & \multicolumn{1}{c}{ } 
      & \multicolumn{2}{c}{765}& 
      \\ 
      \hline
      && exp \footnotemark[4]
      & \multicolumn{1}{c}{($\uparrow$)} 
      & \multicolumn{2}{c}{750}& 
      \\
      \hline
      {\it Imma}$\leftrightarrow$sh 
      & R  & GGA 
      & ($\uparrow$) 
      & 1032 
      & 1029--1031 
      & ($\downarrow$) 
      \\ 
      & HL & GGA 
      & ($\uparrow$) 
      & 1010
      & 965 
      & ($\downarrow$) 
      \\ 
      & EC & GGA 
      & \multicolumn{1}{c}{ } 
      & \multicolumn{2}{c}{1009}& 
      \\ 
      \hline
      && exp \footnotemark[4]
      &\multicolumn{1}{c}{($\uparrow$)} 
      &\multicolumn{2}{c}{$>$ 810}& 
      \\
    \end{tabular}
  \end{ruledtabular}
  \footnotetext[1]{Refs.~\onlinecite{McMahon1993,McMahon1994,Hu1986,Olijnyk1984,Hu1984,Werner1982,Spain1984,Zhao1986}.}
  \footnotetext[2]{Ref.~\onlinecite{McMahon1994}.}
  \footnotetext[3]{Refs.~\onlinecite{Olijnyk1984,Menoni1986,Yoshiasa1997,Werner1982,Spain1984,Polian1990}.}
  \footnotetext[4]{Refs.~\onlinecite{Nelmes1996,McMahon1996}.}
\end{table}

For Si the value of the pressure for the cd$\to${$\beta$}-tin
transition calculated with GGA is within the range of the
experimental values whereas the LDA calculation underestimates
it. We could not reproduce the experimentally observed
hysteresis\cite{McMahon1994} for the $\beta$-tin$\to${\it Imma}
transition in Si.  For a first-order phase transitions, such as {\it
Imma}$\to$sh, a hysteresis could occur due to the behaviour of the
hydrostatic line as in Fig.~\ref{GebcoH}. Since our results support
a second-order phase transition for $\beta$-tin$\to${\it Imma} rather
than a first-order one, no hysteresis should be found in this case.
The transition pressures mentioned in Ref.~\onlinecite{McMahon1994}
are obtained either from the first appearance of a new phase or from
the complete transition of the sample from one phase to the
other. Thus we estimate the experimental range in which the
$\beta$-tin$\to${\it Imma} and {\it Imma}$\to$sh transitions occur
in analogy to the procedure explained in
Section~\ref{pressure}. Unfortunately, the calculated transition
pressures for $\beta$-tin$\to${\it Imma} lie below the ones for the
first cd$\to${$\beta$}-tin phase transition and underestimate the
experimental ones.

The next phase transition to be considered is {\it
Imma}$\to$sh. Here a hysteresis was found within our calculations as
well as in the experiment. The results for the transition pressures
within LDA underestimate again the experimental ones, whereas the
ones within GGA overestimate them. Not included in
Table~\ref{summary} are the experimental pressures for mixed (50:50)
{\it Imma}-sh samples, which are 154~kbar for pressure increase and
159~kbar for pressure decrease\cite{McMahon1994}. This is the only
case in which the transition to the (mixed) {\it Imma} phase occur
at a lower pressure for an increasing than for a decreasing external
pressure. All other experimental points show the same kind of
hysteresis which is reproduced in our results. The difference of the
pressures, where the {\it Imma} phase was measured with pressure
increase and decrease is 5~kbar for the mixed samples and
11--12~kbar for the pure ones. From the experimental transition
pressures the hysteresis is found to be between 3 and
14~kbar. Within our results the largest possible hysteresis is found
to be around 29 (24)~kbar and the smallest around 1--12 (2--7)~kbar
for GGA (LDA), which is in agreement with the experimental
values. The transition pressures determined by the crossing of the
enthalpy curves are found to be within the hysteresis range
calculated here.

For Ge the transition pressure for the cd$\to${$\beta$}-tin
transition is slightly smaller than the experimental one but the
results for the $\beta$-tin$\to${\it Imma} transition match much
better. In the experiment\cite{Nelmes1996,McMahon1996} just the
$\beta$-tin$\to${\it Imma} transition was examined, because the
transition pressure for the {\it Imma}$\to$sh transition was higher
than the maximum accessible pressure of 810~kbar. In comparison to the
experimental results of Si it is assumed that the {\it Imma} phase
for Ge is stable from 750 to 850~kbar, which is compatible with the
observation of a sh phase at 900~kbar.\cite{Vohra1986} From this
point of view, the transition pressure calculated here is slightly
too high.

There are many reasons for the discrepancy between our results and
the experimental ones. On one hand, the experimental conditions like
pressure profiles or measurement times are not perfectly controlled,
whereas nonhydrostaic conditions affect the transition pressures; on
the other hand, temperatue effects, which are neglected in our
calculation, could be of some relevance.  In fact, for Si the
theoretical transition pressure decreases with increasing
temperature.\cite{Gaal2001}
%
%
\section{Conclusions}\label{conclusion}
We have presented a first-principles investigation of the
pressure-induced phase transitions cd $\to$ $\beta$-tin $\to$ {\it
Imma} $\to$ sh in Si and Ge. Numerical accuracy has required these
calculations to be performed within the same unit cell and the
pressure to be calculated from the stress tensor instead of using
any equation of state. We have determined the phase transitions by
investigating the relaxed parameters, the enthalpy vs.\ pressure
curves, and the hydrostatic condition. From the pressure dependence
of the relaxed structure parameters the hysteresis effect does not
seem to have been verified in a theoretical investigation before. An
examination of the hydrostatic condition provides a thermodynamical
explanation for the existence of this hysteresis. Further on, the
orders of the phase transitions have been investigated with the
helps of the curves of the enthalpy vs.\ pressure and of the
derivative of the enthalpy vs.\ pressure. For Si the transitions
{\it Imma}$\to$sh and $\beta$-tin$\to${\it Imma} are found to be of
first and of second order, respectively. For Ge the results support
a second-order $\beta$-tin$\to${\it Imma} transition more than a
first-order one, but in this case the order cannot be definitely 
determined.  A calculation of the enthalpy barriers between the
phases show a vanishing barrier between the $\beta$-tin and the {\it
Imma} phase for Si, which also supports the second-order
$\beta$-tin$\to${\it Imma} transition, whereas no barrier could be
found for all $\beta$-tin$\to${\it Imma}$\to$sh
transitions. Finally, our alternative results are fully consistent
with each other and show an acceptable agreement with the available
experimental data.
%
%
\section*{Acknowledgement}
Support by the Heinrich B\"oll Stiftung, Germany, is gratefully
acknowledged. 
%
%
\appendix
\section{Fermi energy and band structure}\label{App}
\begin{figure}[ht]
  \epsfig{figure=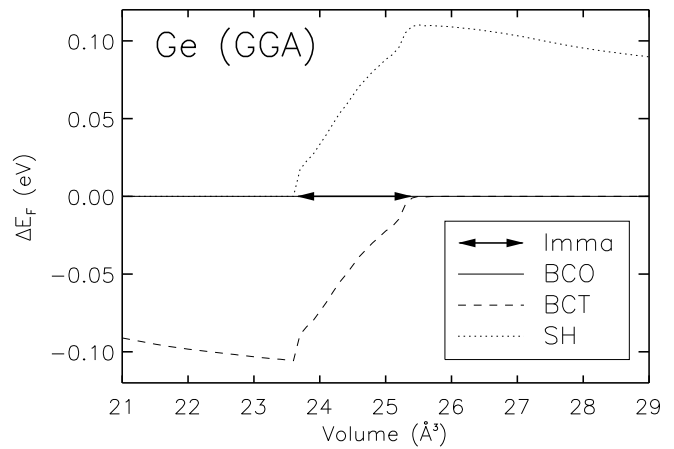, width=8.6cm,angle=0} 
  \caption{Fermi energies of the involved structures for Ge 
    reduced by the Fermi energy of BCO, taken as zero.}\label{fermi}
\end{figure}

We examined the volume dependence of the Fermi energy of BCT, BCO,
and SH. In Fig.~\ref{fermi} we show the Fermi energy reduced by 
the BCO values for Ge,
\begin{eqnarray}
\Delta E_{\rm F}(V) = E_{\rm F}(V)-E_{\rm F}^{\rm BCO}(V) \quad .
\end{eqnarray}

In the range of stability of the $\beta$-tin phase the Fermi energy
of BCO is identical with the one of BCT. Similarly, the Fermi energy
of BCO and SH are identical in the range of stability of the sh
phase. Therefore, the regime where the Fermi energy of BCO is
neither equal to the one of BCT nor to the one of SH is assumed here
to be the range of stability of the pure {\it Imma} phase.

\begin{figure}[ht]
  \epsfig{figure=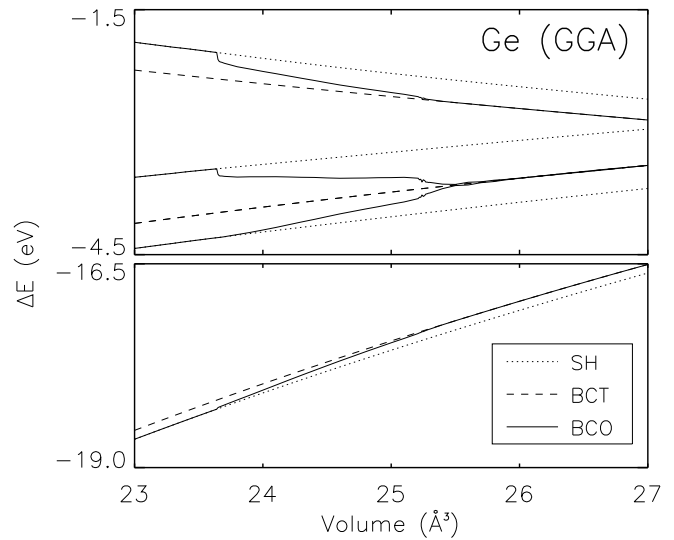, width=8.6cm,angle=0}
  \caption{Eigenenergies at the $\Gamma$ point, reduced by the
    Fermi energy for the BCO, BCT and SH 
    structure for Ge.}\label{eigen}
\end{figure}

We analysed the volume dependence of the band structure for the BCT,
BCO, and SH structure, too. The results for the lowest four bands at
the $\Gamma$-point for Ge are displayed in Fig.~\ref{eigen}. 

The energy values $E_i$ are reduced in such a way that the Fermi
energy at any volume is set equal to zero,
\begin{eqnarray}
\Delta E(V) = E_i(V)-E_{\rm F}(V) \quad .
\end{eqnarray}
In most of the volume range presented in Fig.~\ref{eigen} the band
structure presents a linear dependence on volume. However, the
energy gradients differ for each structure. Within the transition
region between the {\it Imma} and the sh phase one observes a small
discontinuity (here apparent at $V=23.64$\AA$^3$) which is
attributed to the first-order phase transition. When starting from
the $\beta$-tin phase, a deviation of the BCO band energies from the
BCT energies can be observed, which increases continuously. Thereby,
the degeneracy of the BCT bands is lifted, which indicates the
appearance of the less symmetrical {\it Imma} phase. The behavior of
the band structure for Si is very similar to the one of Ge.

%
%

\end{document}